\documentclass[conference]{IEEEtran}
\IEEEoverridecommandlockouts
\usepackage{cite}
\usepackage{amsmath,amssymb,amsfonts}
\usepackage{algorithmic}
\usepackage{graphicx}
\usepackage{textcomp}
\usepackage{xcolor}

\usepackage{float}
\usepackage{caption}
\usepackage{subcaption}
\usepackage{hyperref}
\hypersetup{
    colorlinks=true,
    linkcolor=blue,
    citecolor=blue,
    urlcolor=cyan,
}

\usepackage{algorithmic}
\usepackage{algorithm}
\def\BibTeX{{\rm B\kern-.05em{\sc i\kern-.025em b}\kern-.08em
    T\kern-.1667em\lower.7ex\hbox{E}\kern-.125emX}}
\begin{document}
\title{A Serverless Engine for High Energy Physics Distributed Analysis}

\author{
\IEEEauthorblockN{1\textsuperscript{st} Jacek Kuśnierz}
\IEEEauthorblockA{\textit{Institute of Computer Science} \\
\textit{AGH }\\
Kraków, Poland \\
\textit{Department of Informatics, TUM}\\
Munich, Germany \\
kusnierz@protonmail.com}
\and
\IEEEauthorblockN{2\textsuperscript{nd} Vincenzo E. Padulano}
\IEEEauthorblockA{\textit{EP-SFT, CERN}\\
Geneva, Switzerland \\
\textit{DSIC, UPV}\\
Valencia, Spain \\
vincenzo.eduardo.padulano\\@cern.ch}
\and
\IEEEauthorblockN{3\textsuperscript{rd} Maciej Malawski}
\IEEEauthorblockA{\textit{Institute of Computer Science} \\
\textit{AGH }\\
Kraków, Poland\\
malawski@agh.edu.pl}
\and
\IEEEauthorblockN{4\textsuperscript{th} Kamil Burkiewicz}
\IEEEauthorblockA{\textit{Institute of Computer Science} \\
\textit{AGH }\\
Kraków, Poland}
\and
\IEEEauthorblockN{5\textsuperscript{th} Enric Tejedor Saavedra}
\IEEEauthorblockA{\textit{EP-SFT} \\
\textit{CERN}\\
Geneva, Switzerland}
\and
\IEEEauthorblockN{6\textsuperscript{th} Pedro Alonso-Jordá}
\IEEEauthorblockA{\textit{DSIC,UPV}\\
Valencia, Spain \\
palonso@upv.es}
\and
\IEEEauthorblockN{7\textsuperscript{th} Michael Pitt}
\IEEEauthorblockA{\textit{EP-CMG-OS} \\
\textit{CERN}\\
Geneva, Switzerland}
\and
\IEEEauthorblockN{8\textsuperscript{th} Valentina Avati}
\IEEEauthorblockA{\textit{EP-UHC} \\
\textit{CERN}\\
Geneva, Switzerland}
}
\maketitle

\begin{abstract}
The Large Hadron Collider (LHC) at CERN has generated in the last decade an unprecedented volume of data for the High-Energy Physics (HEP) field. Scientific collaborations interested in analysing such data very often require computing power beyond a single machine. This issue has been tackled traditionally by running analyses in distributed environments using stateful, managed batch computing systems. While this approach has been effective so far, current estimates for future computing needs of the field present large scaling challenges. Such a managed approach may not be the only viable way to tackle them and an interesting alternative could be provided by serverless architectures, to enable an even larger scaling potential.

This work describes a novel approach to running real HEP scientific applications through a distributed serverless computing engine. The engine is built upon ROOT, a well-established HEP data analysis software, and distributes its computations to a large pool of concurrent executions on Amazon Web Services Lambda Serverless Platform. Thanks to the developed tool, physicists are able to access datasets stored at CERN (also those that are under restricted access policies) and process it on remote infrastructures outside of their typical environment. The analysis of the serverless functions is monitored at runtime to gather performance metrics, both for data- and computation-intensive workloads.

\end{abstract}


\section{Introduction}
\subsection{High Energy Physics and CERN}

CERN is the largest research centre for HEP, attracting a wide range of scientists and collaborating with various institutions from all over the globe. It hosts the Large Hadron Collider (LHC), a machine which has generated an unprecedented volume of data in the last few decades, reaching spikes of a few PB per year. Managing and processing information about physics events happening in the accelerator has already proven to be a challenge. The last active period of the accelerator saw peaks of TB of data being generated each day. Furthermore, current estimates for future computing and storing needs describe an increase of a factor 30 in data size~\cite{elsen}. Thus, research in this field needs to focus on performant and scalable software to address the challenges ahead~\cite{CMSOfflineComputingResults}.

\subsection{State of the art in High Energy Physics (HEP) data processing}
The full-scale analyses in which large scientific collaborations at CERN and worldwide are interested in require processing massive data volumes. A modern-day single computer, no matter how powerful, is not fit for performing this kind of task within a reasonable time frame. All the same, the scientists need to lower their data processing workflow runtime as much as possible to get to their physics results faster. To this end, distributed computing approaches involving thousands of nodes were developed and put into practice at CERN and research institutions worldwide. 

The most common solution for a HEP scientist to use these pools of resources is provided by batch processing systems such as HTCondor~\cite{HTCondor}, implemented at a large scale in the Worldwide LHC Computing Grid (WLCG)~\cite{wlcg}. This kind of approach comes with a cost: the need to manage the infrastructure, as well as a strict requirement of getting to know its configuration language and preparing the submission file. All of that work is done just to be able to distribute the analysis processing on the cluster machines. This requires some training even for the end user who, while not having to know about the underlying hardware, still has to write the configuration file in a domain-specific language not used anywhere else.

The nature of batch processing systems does not fit well with the goal of giving end users the ability to run their analyses in an interactive environment. This approach has gained increasing popularity in the HEP field in the last few years, also sustained by a similar, more generic trend happening at wide in the industry, which relies on widely used data science methods and tools. CERN has joined this trend with the deployment of an interactive web-based interface for physicists, named SWAN~\cite{SWAN}, that allows running applications in an interactive Jupyter notebook environment~\cite{jupyter}. The system also features an integration with an Apache Spark~\cite{Spark} cluster at CERN, so that more computationally intensive applications can be seamlessly offloaded to remote resources from within the same interface.

While this type of system already abstracts quite a lot of dependencies and configuration for the users, letting them focus mostly just on the operations they need to perform on a dataset, it is still bound to a specific infrastructure deployment. The physical resources and the software stack need to be provided and managed by a central agent.

\subsection{Serverless computing}

Serverless computing is a cloud computing execution model in which the cloud provider allocates machine resources on demand, taking care of the servers on behalf of their customers. It makes it easier for developers to access a high amount of computing power without having to think about setting up or managing the machines executing the code. It operates on the notion of "function" - a piece of code that will be replicated and invoked on multiple nodes.

The introduction of the serverless computing paradigm improved on some of the pitfalls of managed systems described above. For example, the end user does not need to have any knowledge of the underlying cluster infrastructure while submitting tasks to the remote machines. This allows to run massively parallel computations outside of typical supercomputing facilities, as less deployment-specific administration overhead is required. 

The old approaches, such as HTCondor, running on typical HEP environments struggle with scaling to very high numbers of concurrent executions in multiuser settings, limited by the hardware available for the computations. 

Furthermore, they are not as intuitive or user friendly and put an extra burden on the user who needs to split the distributed application in multiple separate steps.

\subsection{State of the art in serverless data processing}

Every serverless execution engine relies on a type of abstraction layer for the infrastructure itself. This is usually provided by big vendors, such as Cloud Functions by Google\cite{gcf}, Lambda by Amazon, or open source solutions such as OpenWhisk~\cite{OpenWhisk} running on real or virtual machines or Knative~\cite{Knative} running ephemeral Docker~\cite{docker} containers directly on Kubernetes~\cite{Kubernetes}.

Efficient orchestrating frameworks are useful in utilizing the power of serverless functions in data processing applications.

One among these would be PyWren~\cite{PyWren}. It allows to seamlessly distribute arbitrary Python code over multiple nodes. Needed objects and dependencies are serialized through \textit{cloudpickle}~\cite{cloudpickle} in order to execute the application on AWS Lambda natively. As of 2021, the original project is no longer maintained, but it was used as a basis for interesting extensions, including NumPyWren for numerical algebra~\cite{numpywren}. The newer development, Wukong~\cite{wukong}, builds on NumPyWren experience with a focus on data locality and improved, decentralized scheduling.

Most serverless data processing frameworks that we are aware of require writing specific configurations for the analysis not only in generic files, but also within the analysis code. That is why here the comparison will be made only with those frameworks that do not require analysis code changes, same as the developed solution.

The serverless research scenario is quite wide. Other frameworks do not feature a transparent offload of the Python code from one machine to multiple functions, but have still proven to be effective in some regards. For example, the implementation given by the SCAR framework has allowed executing arbitrary container environments on AWS Lambda, even before this option was officially supported by the cloud provider~\cite{SCAR}.

\subsection{Goal of the project}
Allowing HEP scientists to benefit from the advantages of serverless computing without needing to teach them completely new tools requires a new method of communicating with existing platforms.

This work presents a serverless engine that has been developed to run previously existing HEP analysis applications with almost the same code in a seamlessly parallelized way with massively distributed serverless functions. We have described an early proof-of-concept of our engine in a short communication~\cite{HPDC_article}. In this paper, we present a comprehensive and complete solution, with the following main contributions:
\begin{itemize}
    \item new scalable serverless approach for processing data from HEP based on RDataFrame extension of ROOT framework,
    \item implementation of the scalable distributed processing engine which allows running dynamically compiled C++ code on AWS Lambda;
    \item solution to the problem of accessing private experiment data from the cloud based on transferable Kerberos tickets;
    \item detailed performance evaluation using a real physics analysis of PPS project of the CMS experiment from CERN and a synthetic CPU-intensive benchmark.
\end{itemize}

The paper is organized as follows: in Section~\ref{tools} the parts used in the engine are described. In Section~\ref{engine} architecture and algorithms of the engine are presented. Section~\ref{experiments} contains two different analyses that the engine was benchmarked with, as well as the results achieved through them. Section~\ref{conclusions} discusses achievements and future work. Section~\ref{ack} ends the work with thanks to the patrons of this paper. 

\section{Tools}
\label{tools}
As mentioned in Goals, our prototype processing engine uses AWS Lambda as the execution environment. ROOT is the framework in which the analysis is written. XRootD communicating with EOS enables data access by analysis on AWS, authorized via Kerberos ticket. The last part of the developed solution is Terraform, which automates the deployment of all parts of the infrastructure running the computations. These tools are presented and discussed below in the scope used in the project. 

\subsection{AWS Lambda}
AWS Lambda \cite{AWSLambda} is a service that allows running event-driven short-lived computations in a serverless environment. The platform provides a user-friendly way of deploying user's code in a broad range of languages to be run on the cloud provider infrastructure. With the rising popularity of containerization technologies such as Docker, the AWS introduced support for container-based lambdas as well. When a lambda function signature is present, the execution of it can be triggered when, e.g., a predefined event occurs like uploading a file to the AWS data storage or sending an explicit HTTP request. The key feature is that an administration of underlying resources is solely done by AWS and one can dedicate time to perfect the application logic instead of managing servers or virtual machines. Another thing that is worth mentioning is the high scalability of this approach. Once deployed, scaling is done automatically on demand, which means that the user can easily call for more computing power at any time.

The important thing to note here is that once the container is spawned for function execution, it is retained for around 30 minutes, depending on the current workload on AWS datacentre~\cite{lambda_env}. Further invocations running on the same container will not have to wait until the whole environment has been prepared - meaning that reuse of the container will have made the startup shorter by that time. When reusing the existing in-memory container, we are dealing with much shorter ``warm starts'' compared to initializing from scratch ``cold starts'' of Lambda functions.

\subsection{ROOT}

ROOT~\cite{ROOT} is the most widely used software framework for storing, analysing, processing, and displaying HEP data. It has seen wide adoption at CERN and several other institutions worldwide connected with it, such as those participating in WLCG.

The framework defines a common data structure and data layout to store HEP datasets, called TTree~\cite{ttree-class}. Its layout is columnar on the disk, so that different columns can be treated independently. The ROOT I/O subsystem is able to read just a portion of a dataset, to minimize read requests to the filesystem. The minimal amount of information that can be read independently from other parts of the file is called a cluster, which corresponds to a range of entries that can belong to one or more columns. ROOT datasets can be stored and read within the local filesystem of the user machine, but very often are located in remote, distributed storage systems and can be accessed through remote protocols like HTTP or XRootD.

The main interface for analysing a TTree (and other data formats) within ROOT is called RDataFrame~\cite{RDataFrame}. With RDataFrame, users can focus on their analysis as a sequence of operations to be performed on the dataset, while the framework takes care of the management of the loop entries as well as low-level details such as IO operations and parallelization, effectively creating a computation graph, that is a directed graph where each node corresponds to one of the operations to be performed on data. RDataFrame provides methods to perform most common operations required by HEP analyses, such as \texttt{Define} to create a new column in the dataset or \texttt{Histo1D} to create a histogram out of a set of values. Other than TTree, the interface supports processing datasets stored in formats like CSV, Apache Arrow or NumPy arrays. Users can also create an empty dataset with a certain amount of rows that can be filled through the operations in the API. This is particularly useful for benchmark and simulation scenarios.

RDataFrame has been built with parallelism in mind. In fact, it is natively able to exploit all cores of a single machine through the implicit multithreading interface available in ROOT.

\subsection{Kerberos}
MIT Kerberos\cite{Kerberos} is an authentication protocol which uses passwordless tickets to authorize to a particular entity. After creating and validating a ticket, a client can use it to connect to the server that has issued it. The ticket can be made transferable, meaning that any machine, not only the one that received it, can authorize to the server. Thanks to that, it is possible to authorize any computing node if only such a valid transferable ticket is provided to it.

\subsection{EOS}
EOS~\cite{EOS} is the storage used at CERN. It is a system that the engine interacts with when requesting remote files, if such are declared by the user for the particular analysis.

\subsection{XRootD}
XRootD program \cite{XRootD}, developed by Stanford, is used for high performance, scalable fault tolerant access to data repositories. It allows for redirection and pulling of the data from any authorized instance holding it. In the case of ROOT, it is used to access remote data, usually on EOS enabled servers. It can authorize to them in multiple ways, among them using the Kerberos tickets.

\subsection{Terraform and Infrastructure as Code}
HashiCorp Terraform~\cite{HashiCorpTerraform} is one of many tools allowing declarative management of the infrastructure, also known as Infrastructure as Code. In the case of this project, it was used to automate the process of deployment of the required components, which collectively constitute the computation platform that is used by the engine. An important strength for the HEP scientist is that the connection to the infrastructure requires very little configuration from the user side, just requiring the credentials to AWS to be provided.

After the analysis is done, the user can easily delete the entire infrastructure to ensure no more costs are incurred because of its further idle operation, which would consist only of a few empty storage buckets, a docker image and a function signature.

\section{A serverless engine for HEP distributed analysis}
\label{engine}

\begin{figure}[tbp]
  \centering
  \includegraphics[width=\linewidth]{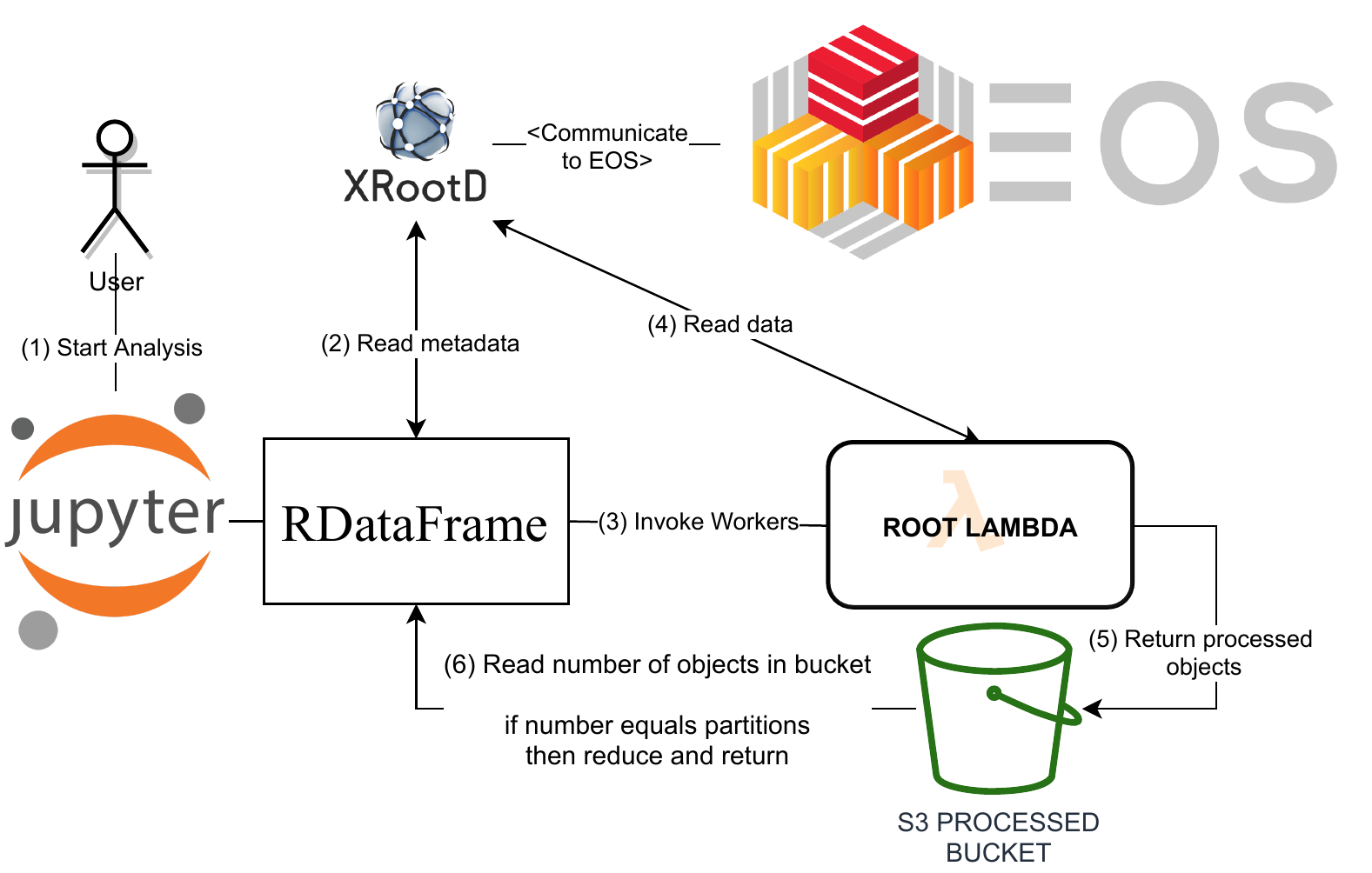}
  \caption{Overview of system's modules including RDataFrame client side, XRootD protocol for communication with EOS data source and AWS Lambda serverless platform and S3 for server side. Engine workflow's sequence described in (n) labels.}
  \label{fig:SimplifiedArch}
\end{figure}

\subsection{Overview}
The goal of the project is to create a distributed execution engine that is strongly integrated with the user-facing analysis framework and does not require almost any configuration. The entire execution from the client perspective is the same as if it is run on the client's  own machine, but underneath it uses distributed computing nodes in the form of serverless functions running on Docker containers. 

The engine relies on the distributed RDataFrame Python package~\cite{DistributedRDataFrame}. This is an extension of the RDataFrame interface that wraps the computations issued by the user in their application code in a MapReduce\cite{mapreduce} pattern. The tool creates tasks that split the dataset of the analysis in various logical ranges of entries. The splitting is done in such a way that the ranges do not overlap and will contain one or more TTree clusters if the original dataset is stored in that format. This ensures that the distributed nodes read the needed entries exactly once. Each range is assigned to a separate serverless function execution that applies the RDataFrame computations to that piece of the original dataset. In this work, the payload of each Lambda execution contains both a range and a copy of the mapper function defined within RDataFrame. When the Lambda is invoked, it executes the mapper on the part of the dataset described by the metadata contained in the range.

The expected input from the user boils down to:
\begin{itemize}
    \item A handle to the dataset: this usually comes in the form of a pair of \texttt{(dataset name, dataset path)}, where the path to the dataset can be provided as a single file path or multiple paths to various files. The path format can change to a URI in case the files are stored remotely.
    \item The Mapper Script: The operations to be processed on the dataset. This is the core of the analysis application and it is written in Python and C++ using the RDataFrame API.
    \item Access token (optional): should the analysis need to access private data stored at CERN, the user has to create and point to the transferable Kerberos ticket to allow the authorization and therefore usage of these resources from both client and worker sides.
    \item Extra C++ code (optional): some more complex physics analyses use additional C++ headers that need to be distributed to the worker environment and declared to the ROOT interpreter to be available at runtime.
\end{itemize}

The engine itself is strongly decoupled: there are separate client and Lambda (also known as worker) sides. The client side is defined by the user application and the machine where it is started. This application uses the correct RDataFrame implementation that is aware of how to connect to AWS and launch multiple Lambda functions. The worker side is managed using Hashicorp Terraform \cite{HashiCorpTerraform} mentioned beforehand, making the deployment of the required infrastructure easy for end user, while also allowing for swift changes or even a complete teardown of the infrastructure by a few commands.

The whole architecture can be seen in Fig.~\ref{fig:SimplifiedArch}. The leading idea was the possibility of changing the backend seamlessly: the AWS Lambda can be exchanged for any other computing service, as long as it can run the environment and allow for temporary storage.

For simplification, the client RDataFrame operating under the newly developed AWS engine attached to ROOT shall be henceforth treated as the client side, and ROOT Lambda as the worker side.

The whole execution environment of the Lambda function is packaged in a container. The reason for this comes from the need of including ROOT, which is too big for typical limits on storage imposed by serverless engines for normal functions. Taking AWS Lambda as an example, the max function package size at the time of writing this paper is 50 megabytes, while a full ROOT installation can reach multiple gigabytes in size.

\subsubsection{Client Side}

The client side extends the RDataFrame, receiving from it both a list of ranges and the user code processed into a single Python object. For creation of the Proof of Concept, the AWS Cloud was used for the serverless platform, although it is possible to change the platform by changing just a few integrating calls. The client side integrates with the AWS using the Boto3~\cite{Boto3} software development kit. The overall mechanism of the client side is described in Algorithm \ref{algo:client}. 

The noteworthy part is that the C++ code from the analysis is packed into a generic Python object using cloudpickle, so it is possible to port the required part of the environment of the client to the worker, allowing for more elastic execution.

The client side also takes care about packaging any required C++ code besides the explicitly declared operations, be it source files or headers. They are sent serialized as an input to the worker main function, which then deserializes and includes them before executing the script from the payload.
\begin{algorithm}
\caption{Engine Algorithm on the Client Side}\label{algo:client}
\begin{algorithmic}[1]
\REQUIRE $npartitions$
\STATE $npartitions \gets int$

\STATE $script \gets$ RDataFrame operations as a single object
\STATE $token \gets$ Read Kerberos auth file
\STATE $data \gets$ Read metadata about declared data sources
\STATE $headers \gets$ Read headers declared in ROOT

\STATE $ranges \gets$ split $data$ into $npartitions$ partitions

\FOR {$range$ in $ranges$}
    \STATE $begin$ ASYNC THREAD($range$)
    \STATE $payload \gets \{range,script,token,headers\}$
    \STATE $\{single\_result, monitoring\_data\} \gets$ call Worker($payload$)
    \STATE call save($monitoring\_data$)
    \STATE $end$ ASYNC
\ENDFOR

\STATE $results \gets$ call reduce($single\_result$) foreach THREAD($range$) 
\RETURN $results$
\end{algorithmic}
\end{algorithm}
\subsubsection{Worker Side}

The worker is provided as a Docker image, utilizing the newly added Docker integration to AWS Lambda~\cite{docker_lambda}. It comes with ROOT installed as well as a customized addition of a monitoring tool. It operates under the Algorithm \ref{algo:worker}, which describes what happens in this particular serverless function. It does not depend on any AWS services with an exception of the S3, allowing for an easy potential portability to other serverless platforms.

When the function is called, it expects to receive the data handle and function to be applied. Optionally, Kerberos tickets and supporting C++ code can be provided. It compiles the C++ part using Cling \cite{Cling} JIT compiler  and executes it on the provided range, downloading it on demand. It is worth noting that Cling underneath makes use of LLVM \cite{LLVM:CGO04}. Therefore, it allows for compiling arbitrary code on the fly, making for a generic worker able to compile and execute any C++ code provided. 

For the measurements for the paper, the customized Python inspector from SAAF~\cite{SAAF} project is employed. It measures the current metrics of the worker environment, including CPU, memory, and bytes transmitted via network. It does so every second, to give a clear and granular understanding of what goes on inside of every single execution to provide more insights for future work. Besides measuring much more often than the original and being run in a separate Python process, it behaves the same as in the SAAF paper. It introduces some overhead, but it is hard to measure the exact influence if the monitoring tool is disabled.

In the current form, no caching is included, although it is possible to attain with several data source string manipulations.

\begin{algorithm}
\caption{Engine Algorithm on Worker Side}\label{algo:worker}
\begin{algorithmic}[1]

\REQUIRE $TokenFilePath$
\STATE $\{range,script,token,headers\} \gets payload$ 
\STATE write $token$ into $TokenFilePath$
\FOR{\{header, header\_file\_path\} in headers}
    \STATE write $header$ into $header\_file\_path$
    \STATE declare $header\_file\_path$ in ROOT
\ENDFOR
\STATE $monitoring \gets$ start ASYNC monitoring process
\STATE $result \gets$ call $script$($range$)
\STATE write $result$ into s3 bucket
\STATE $monitoring\_result \gets$ stop($monitoring$)
\RETURN $monitoring\_result$
\end{algorithmic}
\end{algorithm}

\subsection{Payload}

The payload that is input to the worker function consists of several strings containing serialized objects, such as: 
\begin{enumerate}
    \item Mapper Script;
    \item Range;
    \item Kerberos Ticket;
    \item List of headers, libraries etc.
\end{enumerate}
Aside from Range, the other three objects are identical in all invocations. Kerberos Ticket and headers are serialized on the client side and their contents unpacked to files in temporary storage on the server side.

\subsection{Controlling the execution - asynchronous threads}

Because of the requirement of advanced error handling, the decision was made to use so-called synchronous Lambda invocations. AWS Lambda provides two kinds of Lambda invocations - synchronous and asynchronous~\cite{lambda_invocation}. The asynchronous one is based on the fire and forget mechanism and the response to the invocation request is sent straight away when Lambda queues the event for processing. In contrast to the previous one, the synchronous invocation waits until the computation is finished and reports the execution state directly to the caller. That allows for a more precise retry mechanism, the logic of which is present on the client side. That option is not available in asynchronous calls and would be not trivial to implement. After sending the synchronous invocation request, a connection with the Lambda is established and a thread calling an invocation procedure is waiting for the Lambda to complete processing. When the Lambda's work is done, the thread gets a response containing JSON serialized payload, the content of which can be specified inside the Lambda function's code. This allows for passing information about the success of the computation as well as the errors that occurred during the computation.

Error handling mechanism was achieved by running a pool of asynchronous Python threads, each managing a single synchronous Lambda execution. If the Lambda fails, it sends a type of error and an error message to the client side, where those are logged and a retrial is done on the same thread. A single Lambda instance that still errs after several retrials fails the entire run.

Compared to our initial implementation of asynchronous Lambda invocations on the main Python thread~\cite{HPDC_article}, there was a massive gain in the speed of invocation observed in our experiments, throttled only by AWS documented limit of 10 new synchronous lambda invocations per second~\cite{lambdaquotas} and the speed of connection between client and the AWS server.

\subsection{Kerberos emplacement}

To enable authentication and authorization to secure storage at CERN, a transferable Kerberos ticket needs to be provided by the user, and has to be placed in a location readable by the XRootD on the client side. Once the analysis code is triggered on the client side, it is first used to access the metadata of the required input data files. Then, the contents of the ticket file are serialized and transported in the payload to each worker, which then unpacks it in a specified location for the XRootD to use, allowing for access with the same permissions as on the client side (see Algorithm \ref{algo:worker}). For XRootD to read the ticket, the pointer to it has to be provided. There are several ways to do that. In this implementation, the value of \textit{KRB5CCNAME} environment variable is set on signature of the lambda and used as the target path of writing the ticket file contents from the payload, as well as the path of the file to be read by the XRootD.

\section{Experiments}
\label{experiments}

The goals of the experiments are as follows:
\begin{itemize}
    \item to evaluate the performance and scalability of the engine using two representative workloads;
    \item to collect detailed metrics regarding CPU and network usage, to understand the potential limitations and bottlenecks;
    \item to assess the variability of the underlying AWS Lambda infrastructure under this workload.
\end{itemize}
\subsection{Methodology}
\label{methodology}

The configuration for the a single Lambda function used to run the tests is as follows:
\begin{itemize}
    \item 1769 megabytes of RAM (corresponding to one vCPU-second of credit per second, which means that a full single vCPU core is allocated to this Lambda);
    \item The analysis runs on "Intel(R) Xeon(R) Processor @ 2.50GHz", as reported by monitoring tool;
    \item 15 minutes timeout, which is max value allowed by AWS in time of writing;
    \item Based on a Docker image of size $\thicksim$4 gigabytes~\cite{root_lambda};
    \item AWS Lambda is kept warm by running a 1000 container warmup before the main analysis, meaning that the ROOT initialization time is excluded thanks to avoiding additional time at cold start.
\end{itemize}

Two analyses are run on the infrastructure:
\begin{enumerate}
    \item A CPU-bound benchmark. This serves as a baseline for local resources utilization, as opposed to the typical data-intensive analysis. Creates a simulated dataset with one billion randomly generated entries and three columns storing scalar floating-point values. The simulated dataset total size is 96 gigabytes. The application computes the mean value of each column, ten times per column. This application will be called ``CPU-bound benchmark'' from now on. 
    \item A real physics analysis processing data from the PPS subsystem of the CMS experiment at CERN~\cite{PPS_ANALYSIS}. It consists of the selection of candidate events of exclusive dilepton production, $pp \to p\oplus\ell\ell\oplus p$, with $\ell\in\{ e,\mu,\tau \} $. The measurement of exclusive production of lepton pairs involves  two selections: (1) Exclusive cuts - leptons are produced exclusively, i.e., no other particles are produced during the proton-proton interaction, and (2) correlation between leptons and protons.
    The total dataset size is 420 Gigabytes. This application will be called ``PPS analysis'' from now on.
    
\end{enumerate}

\begin{table}[tb]
  \caption{Comparison of analyses used for testing the created engine.}
  \label{tab:dataset}
  \resizebox{\linewidth}{!}{%
  \begin{tabular}{lll}
    Name&Size~[GB]&Chunks\\
    \hline
    CPU-bound benchmark & 96 & $4\cdot1e9$\\
    PPS analysis & 420 & 26410  \\
\end{tabular}
}
\end{table}

\begin{figure}[!t]
\centering
\begin{subfigure}[t]{0.48\linewidth}
	\includegraphics[width=\linewidth]{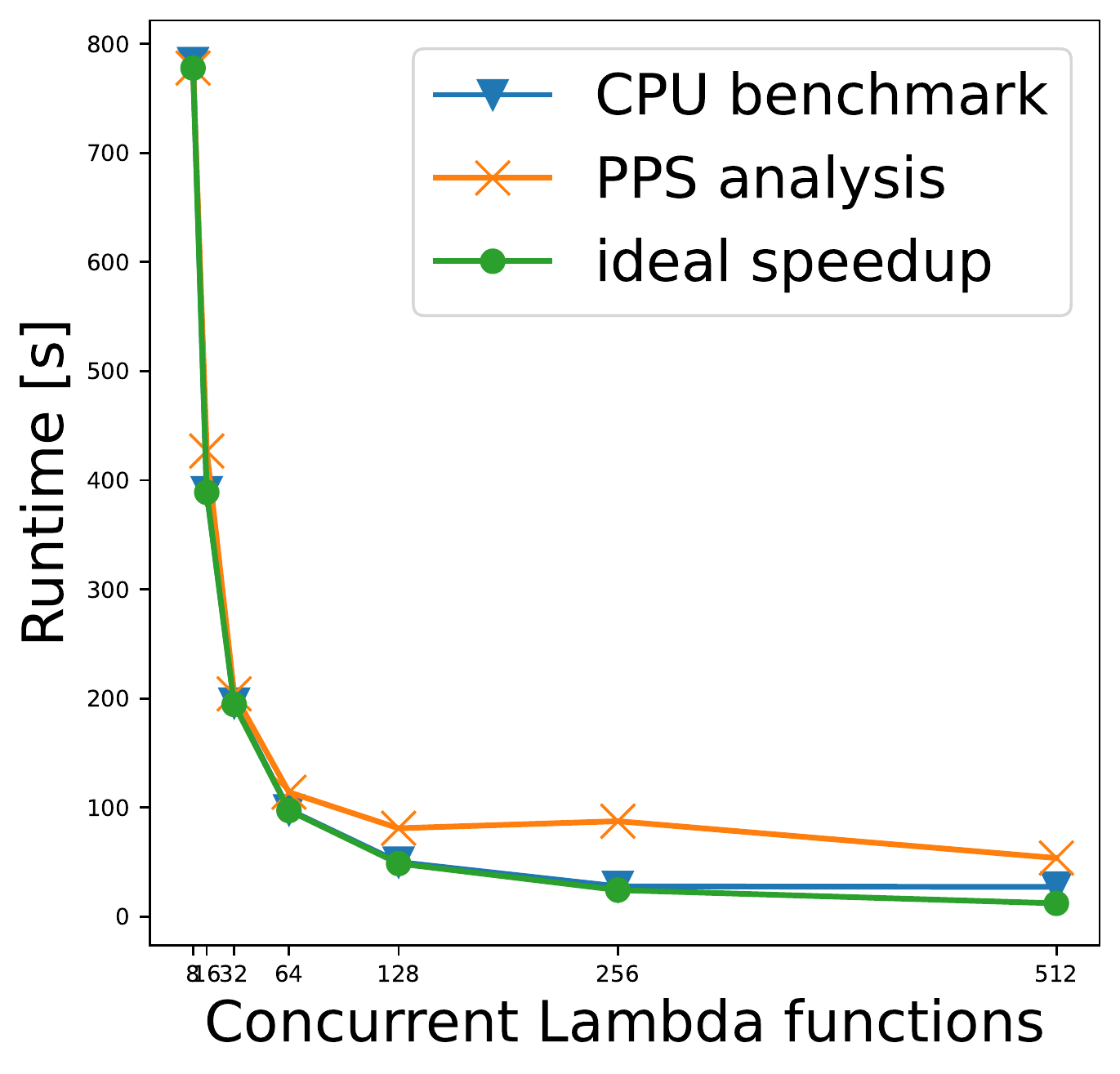}
	\caption{} \label{comparison-runtime}
\end{subfigure}
\begin{subfigure}[t]{0.48\linewidth}
	\includegraphics[width=\linewidth]{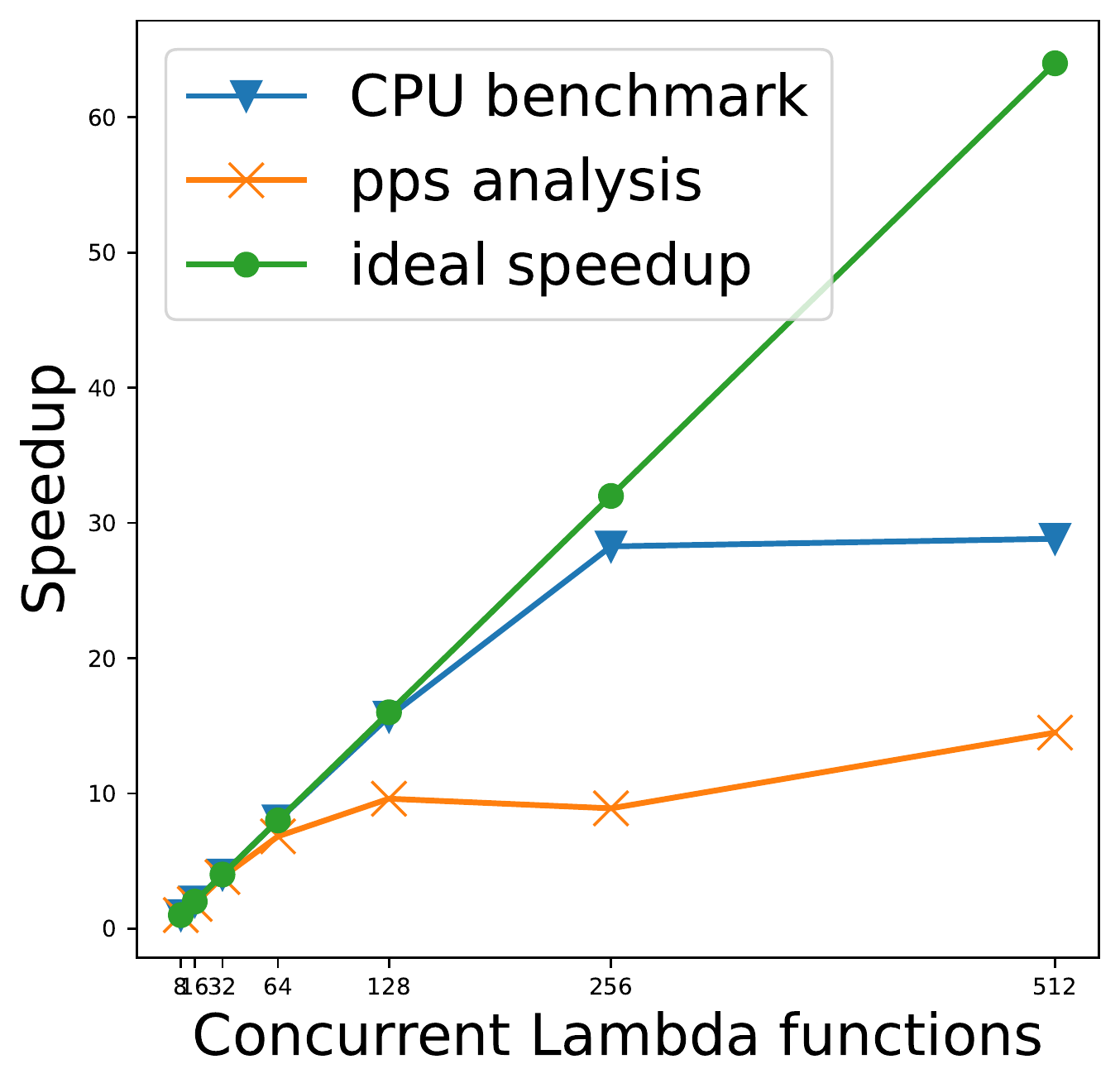}
	\caption{} \label{comparison-speedup}
\end{subfigure}
\caption{Comparison of runtime scaling of the cpu-bound benchmark and the PPS analysis, with an increasing number of Lambda invocations. \subref{comparison-runtime}: comparison of the absolute runtimes. \subref{comparison-speedup}: comparison of the speedup with respect to a linear increase.}
\label{fig:cpu-pps-comparison}
\end{figure}

The datasets' sizes were adjusted to execute just below the Lambda-imposed limit of 15 minutes before timeout (Table \ref{tab:dataset}).

Each analysis is run on a distributed RDataFrame increasing the number of partitions, which is equal to the number of Lambda invocations. These are: $[8,16,32, 64, 128, 256, 512]$. Each invocation processes a separate range of entries of the RDataFrame. The distributed RDataFrame applications are started within an environment that has both the AWS credentials and the Kerberos ticket available (the latter is needed to access the dataset of the PPS analysis).

\subsection{Results}
\label{results}

\subsubsection{Runtime scaling}

The developed engine is put to test with the two analyses described in Section~\ref{methodology}. The total runtime of the execution of the Lambda functions is measured by subtracting the minimum starting time among all executions belonging to a single test run from the maximum ending time within the same run. Figure~\ref{fig:cpu-pps-comparison} shows the behaviour of the system with both applications and an increasing number of Lambda invocations. It should be noted that this figure reports a linear increase in speedup as the ideal expectation. This is due to the specific nature of HEP data. Given a certain collision between particles, all events involved are statistically independent. The logic of an analysis is applied per event and the operations do not change whether there are just one hundred events or billions of them in the same dataset. This means that effectively HEP data analysis is an embarassingly parallel problem. Thus, if no other technical issues were to be encountered, scaling to more cores should follow a linear increase. Such behaviour is also shown in similar works~\cite{DistributedRDataFrame, avati-europar, padulano-acat-2021}. Furthermore, the scalability of the tests in this work is subject to specific limitations imposed by AWS, which are highlighted in Section~\ref{real-utilization-aws-lambda}.

\subsubsection{Network and CPU usage patterns during Lambda execution}

\begin{figure}[htbp]
\centering
\includegraphics[width=\linewidth]{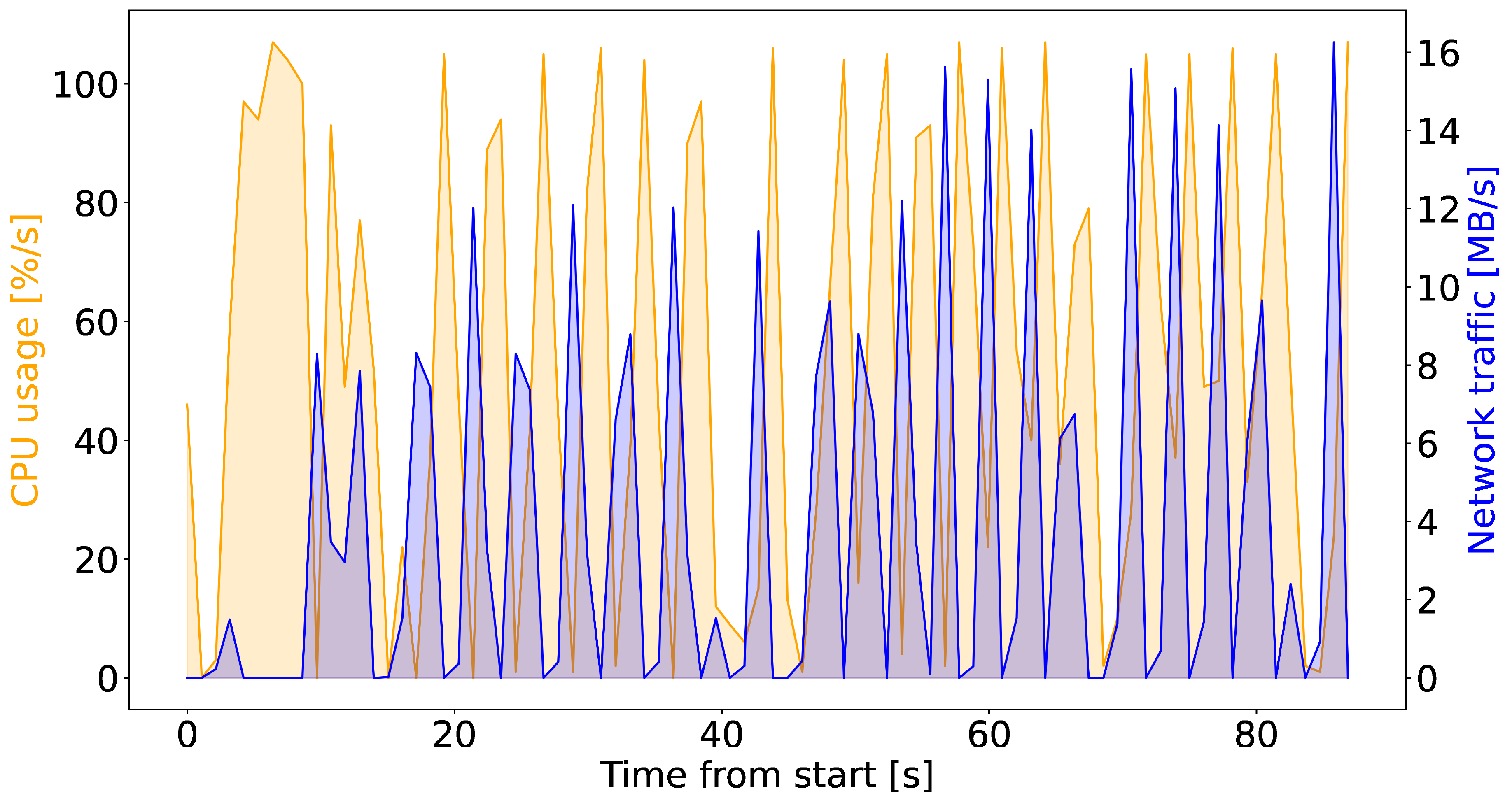}
\caption{CPU usage (in orange) and network traffic (in blue) in a single Lambda execution running PPS analysis. This execution belongs to a run with 64 concurrent invocations.}
\label{fig:networkcpu}
\end{figure}

Running the PPS analysis with the Lambda infrastructure requires streaming the pieces of the dataset needed for the analysis to the functions during their runtime. Given that a certain RDataFrame range can span one or more TTree clusters, these will need to be downloaded by the Lambda when they are needed for processing. The ROOT I/O streams a cluster of entries and processes them as they arrive, leading to the usage patterns shown in Figure~\ref{fig:networkcpu}. This figure focuses on a single Lambda invocation of the whole PPS analysis. During runtime, RDataFrame requests clusters of entries, thus triggering remote read requests that translate to higher spikes in network usage (in blue in the figure). Subsequently, the entries are processed leading to higher CPU usage (in orange in the figure). It can be seen that the spikes in the network and CPU usage alternate one another.

\begin{figure}[htbp]
\centering
\includegraphics[width=\linewidth]{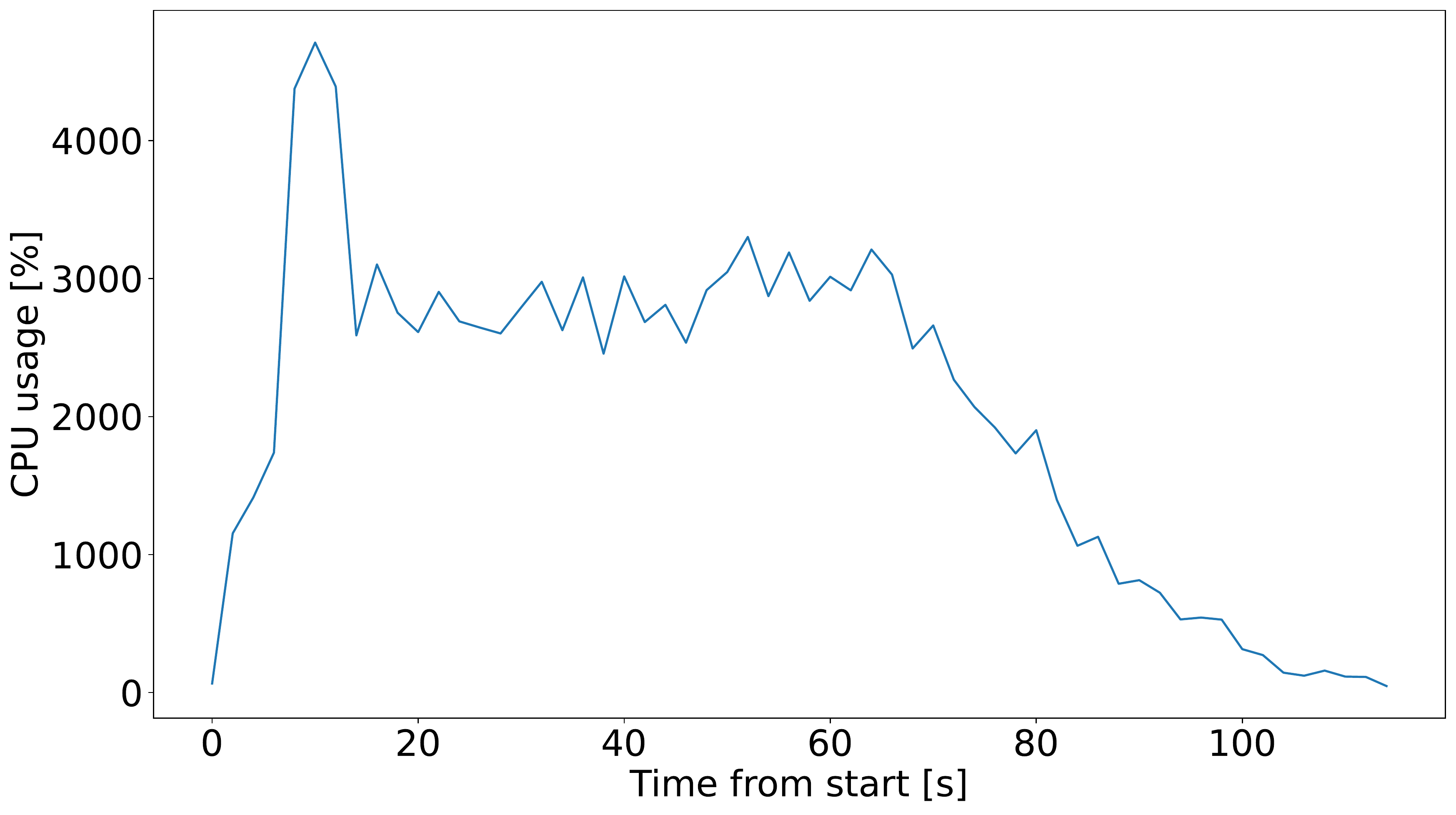}
\caption{CPU usage of 64 Lambda invocations running the PPS analysis.}
\label{fig:cpupercentPPS}
\end{figure}

Aggregating the CPU percentage usage over all Lambda invocations of a particular PPS analysis run produces the pattern shown in Figure~\ref{fig:cpupercentPPS}. Initially, as all lambdas finish downloading the first chunk of the dataset they need to process, a high spike in CPU usage is shown, reaching 74\% of the total available 64 vCPUs. Consecutively, the CPU usage hovers around 40\% until the first Lambda execution finishes and there is a corresponding drop that lasts until all functions finish their workload.

\begin{figure}[htbp]
\centering
\includegraphics[width=\linewidth]{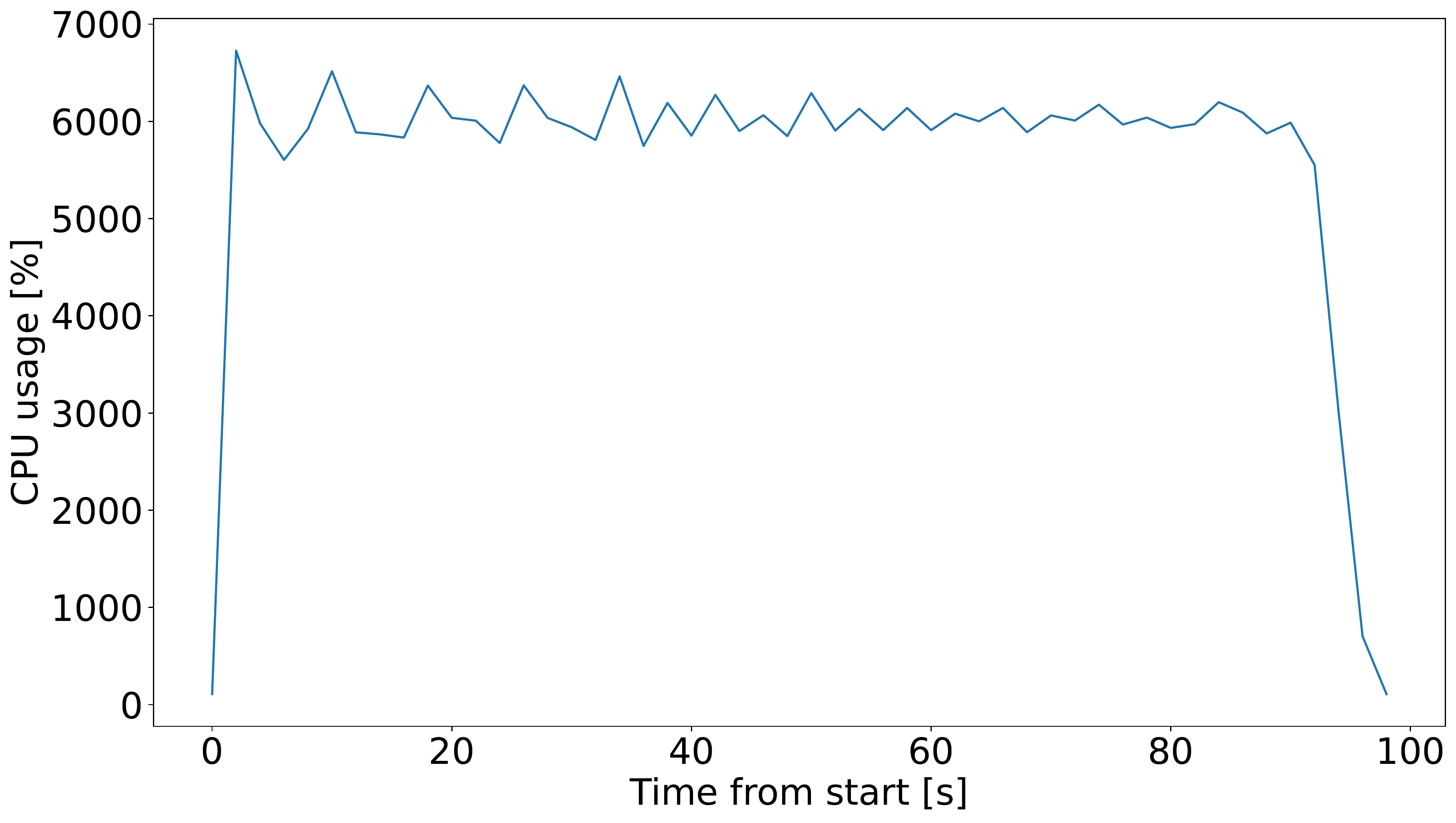}
\caption{CPU usage of 64 Lambda invocations running the CPU-bound benchmark.}
\label{fig:cpupercentcpubound}
\end{figure}

The CPU-bound benchmark shows a different story. In Figure~\ref{fig:cpupercentcpubound} the usage is always above 90\% of the total vCPUs available until the end of the benchmark. This is coherent with the average utilization for every Lambda seen on Figure~\ref{fig:cpu_boxplot}. 

\subsubsection{Variability in starting and ending time of Lambda executions}

\begin{figure}[htbp]
\begin{subfigure}[b]{.48\linewidth}
\includegraphics[width=\textwidth]{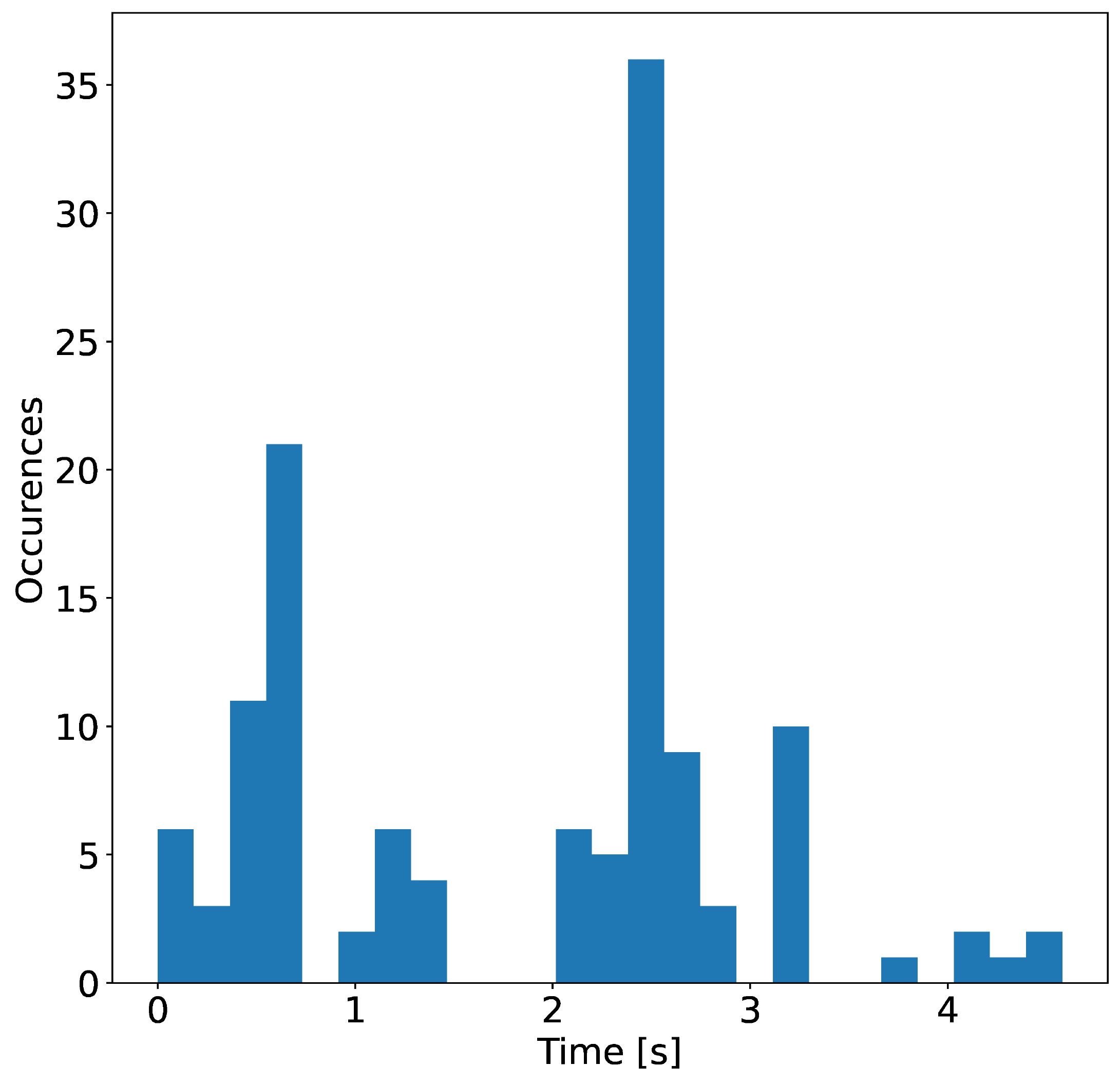}
\caption{Distribution of the start time of a Lambda execution relative to the Lambda that started first.}
\label{fig:relativestarttime}
\end{subfigure}
\hfill
\begin{subfigure}[b]{.48\linewidth}
\includegraphics[width=\textwidth]{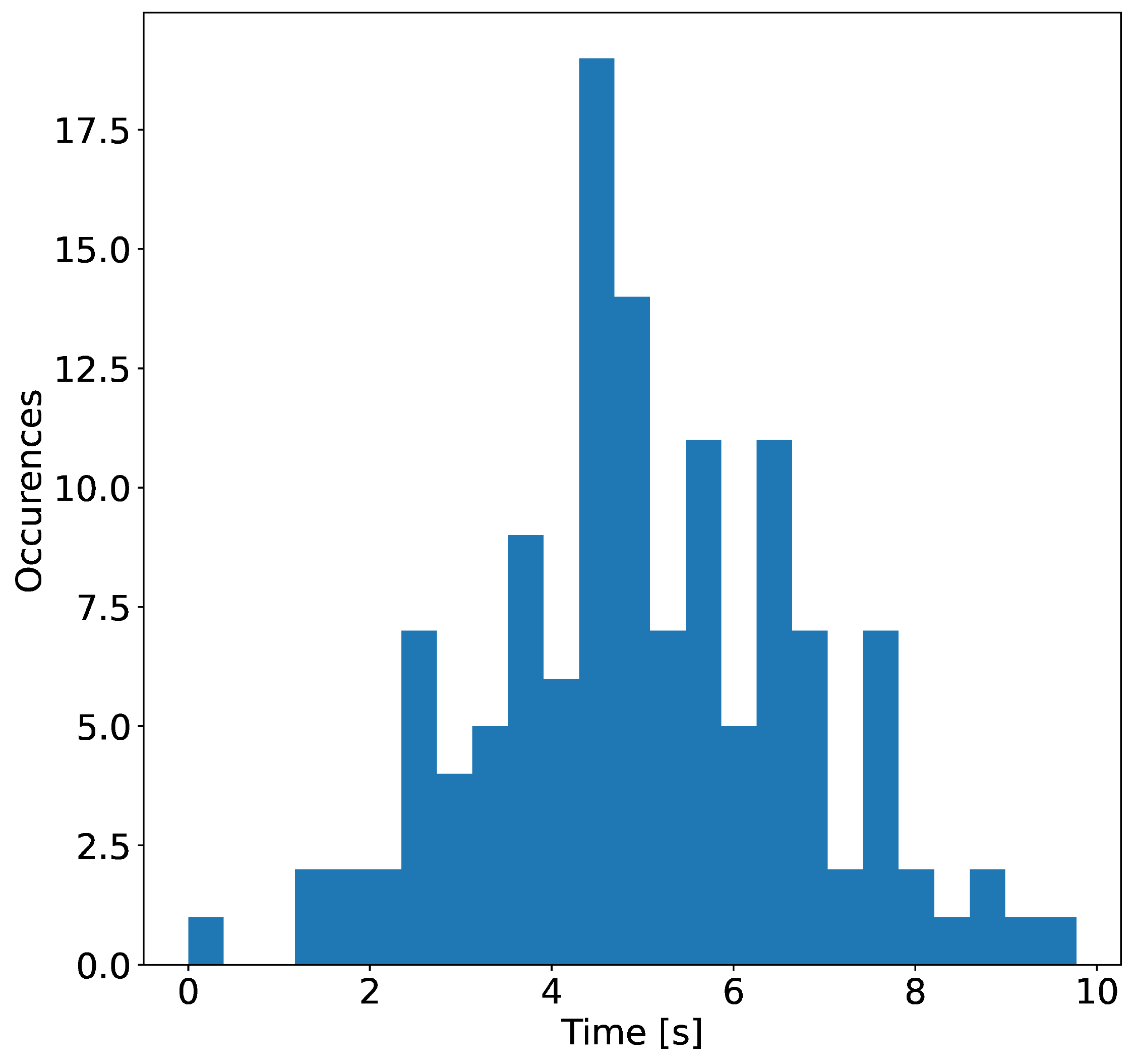}
\caption{Distribution of the end time of a Lambda execution relative to the Lambda that ended first.}
\label{fig:relativeendtime}
\end{subfigure}

\caption{Comparison of distribution for start and end Times for 128 concurrent Lambda invocations running the cpu-bound benchmark. }
\label{fig:relativelambdatimes}
\end{figure}

The monitoring tool running in the Lambda execution also reveals delays in the starting and ending times of different Lambda invocations in the same analysis run. With serverless computing, there is no direct control or access to the computing resources, and this means that the actual starting time of a certain computation after the Lambda has been invoked from the client can show some variability. Figure~\ref{fig:relativelambdatimes} shows that different Lambda invocations in the same test run can vary both in their starting time and in their ending times. In this particular CPU-bound run, the maximum delay between the first starting Lambda and the last one is around 4 seconds, while the maximum delay at the end is around 10 seconds. While quite limited, this delay is still noticeable and it is another peculiarity of the serverless workflow.

\begin{figure}[htbp]
\includegraphics[width=\linewidth]{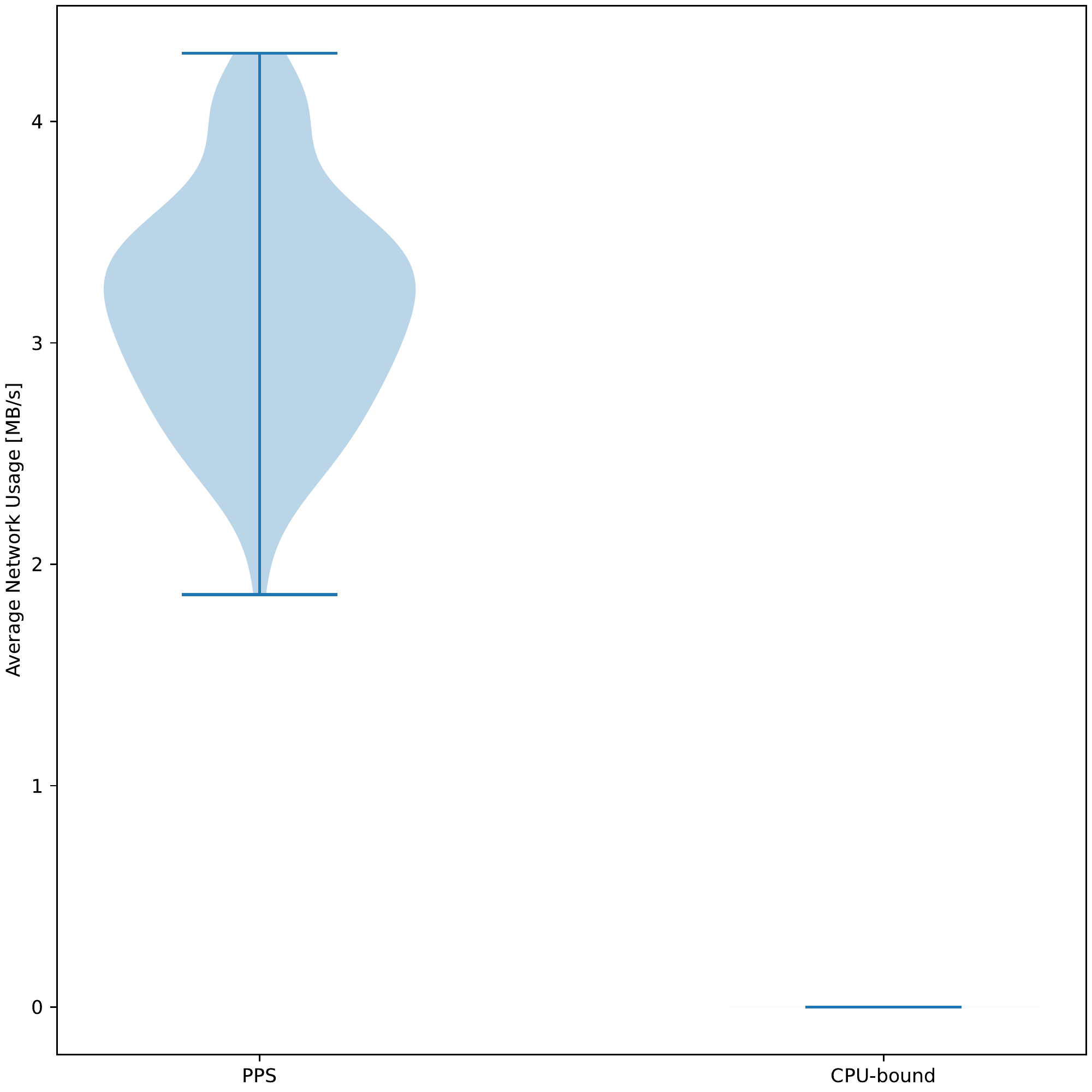}
\caption{Comparison of average network usage for Lambda in both analyses. The CPU-bound is added to show the lack of any network activity.}
\label{fig:network_boxplot}
\end{figure}

\subsubsection{Real utilization of available resources on AWS Lambda}
\label{real-utilization-aws-lambda}

Because of the variable nature of PPS analysis dependant on external IO, the synthetic CPU-bound analysis can be used to see how efficient the serverless function can really be with given resources.

\begin{figure}[htbp]
\includegraphics[width=\linewidth]{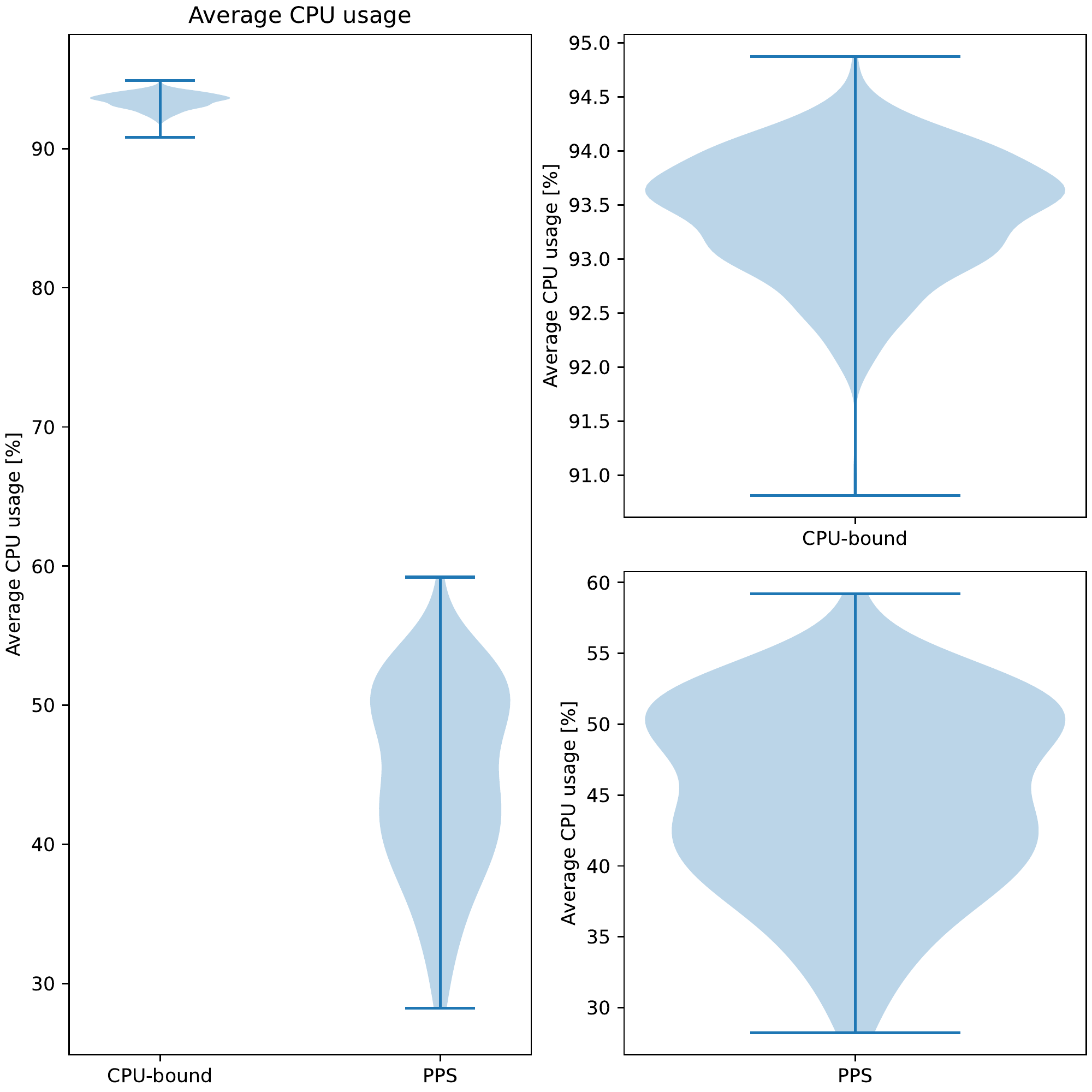}
\caption{Comparison of average CPU usage for every Lambda in both analyses for 512 Lambdas. The left column shows both analyses, while the right shows closeups to both.}
\label{fig:cpu_boxplot}
\end{figure}

We can clearly see that the synthetic CPU-bound benchmark is able to utilize over 90\% of allocated vCPU power at all times (Figure~\ref{fig:cpu_boxplot}). However, the actual time the analysis takes is two times as long as that of the average execution, where the outliers are few and differ little from the average Lambda. This is because of the throttling imposed by AWS on the number of new synchronous invocations per second, leading to delays of startup time for multiple executions.

The PPS benchmark shows greater variability in execution time, with some instances finishing much earlier than the few outliers. Here the overall analysis time is very similar to duration of the longest invocation, but it still takes several seconds longer than it.

\begin{figure}[!t]
\centering
\includegraphics[width=\linewidth]{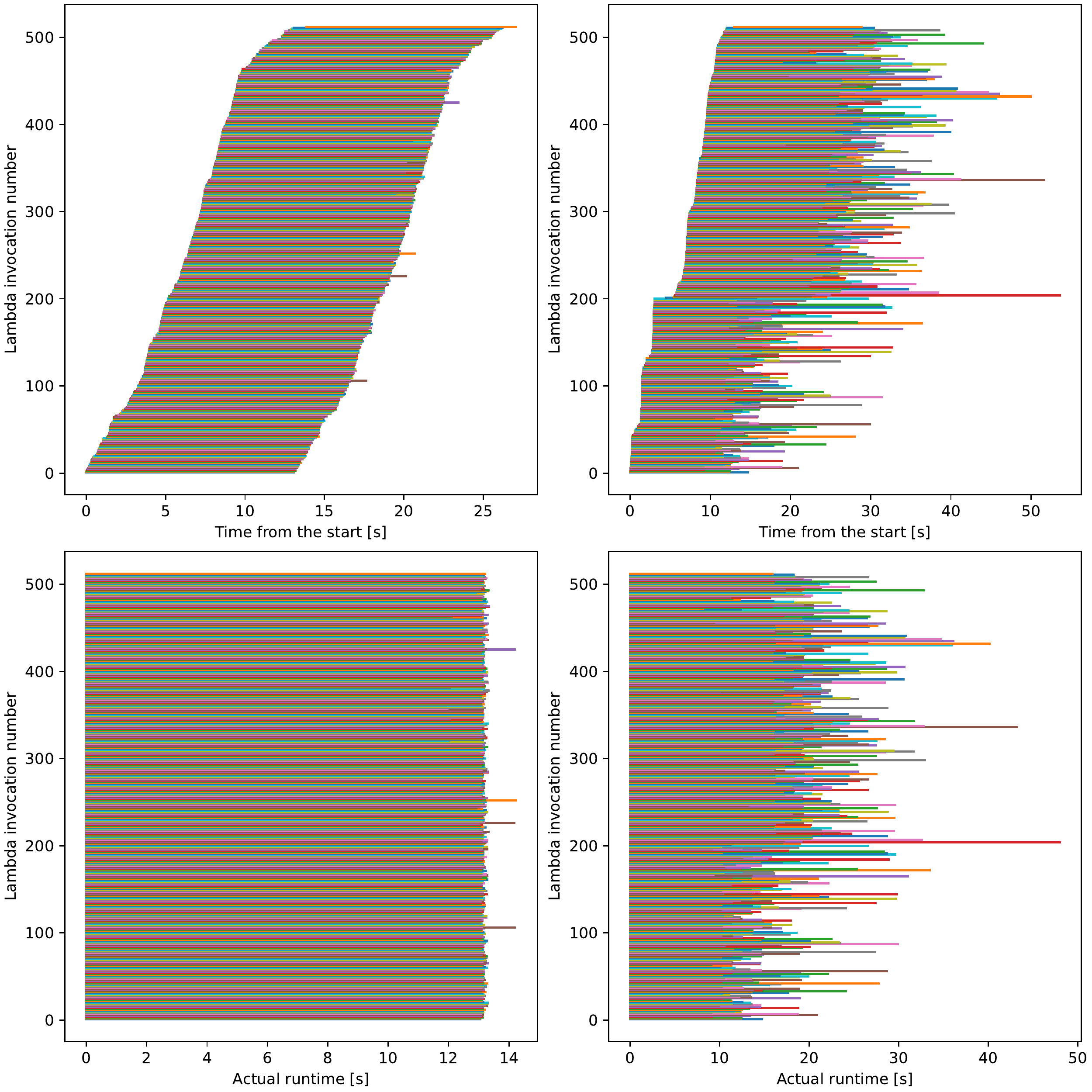}
\caption{Comparison of runtime variability in executions. The top row shows the actual time of a single analysis computation, aligned to the beginning of first Lambda as a 0 point. The bottom row has every Lambdas' start aligned to 0. The left column shows synthetic CPU benchmark, the right PPS analysis.}
\label{fig:cpu-pps-runtime-comparison}
\end{figure}

Both plots on Figure~\ref{fig:cpu-pps-runtime-comparison}  present us with an execution of 512 lambdas given, each in the scope of a single analysis given roughly the same data size. Each line presents a lifetime of a single execution, sorted by start time for a better visibility. 

In the case of PPS analysis, the clear variability is seen. Data is divided as equally as possible, but RDataFrame engine does not have insight into the exact content of the files given to Lambdas and that it can create a difference between the execution times. Moreover, some instances might stumble upon network slowdowns or slower IO with some files that results in a lower utilization of CPU on a particular Lambda. This is why the CPU-bound analysis was taken as a comparison point: we can see that after starting the analysis, the overall execution is very uniform and the outliers are very few. Thus, in principle, RDataFrame is able to saturate the CPU resources available in the Lambda environment.

To understand the results in Figure~\ref{fig:cpu_boxplot}, we need to see the Figure~\ref{fig:network_boxplot} and an example of a single network bound execution on Figure~\ref{fig:networkcpu}. The results show the lack of simultaneous utilization of CPU and network resources.

\subsection{Discussion}

The results shown previously demonstrate how the engine can perform quite well regarding CPU utilization, but there are several limitations imposed by AWS that worsen our current results. The currently implemented synchronous calls to API from outside AWS are severely impacting the startup time, making the execution much longer than it needs to be. 

Despite that, as seen on Figure \ref{fig:cpu-pps-comparison} the scaling shows promises once the limit on the speed of invocation for a higher count of Lambda is solved. While this may seem an issue if only looked at from the perspective of a single user running a single analysis, it is actually a demonstration of how the serverless approach would allow to fully utilize the underlying hardware. In fact, the time periods where the invocations made by the engine have not started yet due to the AWS limit of 10 per second, the same CPUs could still be used by different serverless functions invoked elsewhere. This is a nice compromise between full interactivity and long queues in submission systems usually seen in HPC, and something new to consider for HEP workflows. The other thing is data locality, where the stream of I/O coming directly from CERN causes the PPS analysis to stumble on the network connection between CERN and AWS data centre, as well as probably on CERN storage speed.

The speedup is almost linear up to 128 cores, but above that value it falls off: something that can be fixed by deploying an invocation mechanism on the server side, or using asynchronous mechanisms just like Wukong~\cite{wukong} does. 

\section{Conclusions}
\label{conclusions}

The novelties presented in this work come in many flavors. 

First, this is the earliest example of a public serverless computing platform being used for real HEP analysis used by scientists at CERN, yet it was conducted outside the CERN or affiliated institutions' infrastructure. This was achieved by exploiting a lightweight environment thanks to serverless functions and ephemeral storage to keep track of the current state.

Second, typical serverless frameworks require the user to have some knowledge about distributed computation, which is not required here, as the interfaces in the distributed and local versions are the same. 

Third, this work proves that the current serverless platforms supporting docker are able to run complex analyses on big, complex frameworks requiring a lot of resources even to start up. Scientific platforms are usually limited to particular distributions, or have their functionality limited when using such platforms, but here the full functionality of the distributed RDataFrame is retained.

Fourth, the packaging of all needed software in a single container image allows the user to have a flexible environment, where they can modify the code of even the most complex underlying frameworks such as ROOT and run it on the infrastructure side. This approach was not easy (or sometimes not even possible) on the classical managed infrastructures.

The engine presented here shows a promise for deployment at a large scale as an alternative to a typical on-premises analysis run on grids. The results are promising, as the engine was capable of fully utilizing the AWS resources available during function execution and the only real bottleneck was the limit in Lambda invocations per second. Nonetheless, there is still room for improvement, especially concerning the speed of the overall startup time when running with a large number of concurrent executions, as the network between the client and the AWS API gateway can throttle the initialization part.

In a future work, the engine will be tested and compared with various serverless frameworks and on different compute services, in an attempt to reach the full parallelization capabilities faster. Furthermore, other optimisation techniques will be employed such as including the reduce step in the serverless workflow and exploring input data caching, to further increase performance of the engine.

Another interesting direction to build upon would be the reuse of cling and llvm to run arbitrary C++ regardless of the platform which it is run on. That would result in truly generic C++ script runners, which could be useful for sideloading arbitrary high performance C++ code to Python serverless functions. This could help with the typical binary format of compute-intensive parts of the program, which are not portable between different processor architectures.

\section{Acknowledgements}
\label{ack}
This work was supported by the Polish Ministry of Education and Science, grant DIR/WK/2018/13 and the funds assigned to AGH. It also received support from grant PID2020-113656RB-C22 funded by the Spanish Ministry of Science and Innovation (MCIN/AEI/10.13039/501100011033).

\bibliographystyle{IEEEtran}
\bibliography{references}
\end{document}